\documentclass[aps,10pt,twocolumn,pre,floatfix,longbibliography]{revtex4-2}

\usepackage{graphicx}
\usepackage{amssymb,amsmath,tabularx}
\usepackage{bm}
\usepackage{wrapfig}
\usepackage{hyperref}

\usepackage{xcolor}

\begin{document}
\title{There's Plenty of Room in the Middle:\\ The Unsung Revolution of the Renormalization Group}
\author{Nigel Goldenfeld}
\affiliation{Department of Physics
University of California, San Diego, 9500 Gilman Drive, La Jolla, CA 92093, USA}

\def\Re{\textrm{Re}}
\pacs{}
\begin{abstract}

The remarkable technical contributions of Michael E. Fisher to
statistical physics and the development of the renormalization group
are widely known and deeply influential.  But less well-known is his
early and profound appreciation of the way in which renormalization
group created a revolution in our understanding of how physics --- in
fact, all science --- is practiced, and the concomitant adjustment that
needs to be made to our conception of the purpose and philosophy of
science. In this essay, I attempt to redress this imbalance, with
examples from Fisher's writings and my own work. It is my hope that
this tribute will help remove some of the confusion that surrounds the
scientific usage of minimal models and renormalization group concepts,
as well as their limitations, in the ongoing effort to understand
emergence in complex systems.
\medskip

{\it This paper will be published in \lq\lq 50 years of the
renormalization group", dedicated to the memory of Michael E. Fisher,
edited by Amnon Aharony, Ora Entin-Wohlman, David Huse and Leo
Radzihovsky, World Scientific (in press).}
\end{abstract}
\maketitle

\section{Introduction}

In 1988, Michael E. Fisher authored what must surely be a candidate for
his least cited paper.  Entitled \lq\lq Condensed Matter Physics: Does
Quantum Mechanics Matter?"
\cite{fisher1988condensed,fisher2017condensed} and bestowed with all of
9 citations (according to Google Scholar), it poses an outrageous
question that in Fisher's hands is developed and answered with his
characteristic brilliance, clarity and originality.  Fisher was writing
at the behest of Herman Feshbach to review the state of condensed
matter physics as it stood on the 100th anniversary of the birth of
Niels Bohr, with a special remit to comment on the overlap with Bohr's
ideas on the fundamentals of quantum mechanics.  The timing could not
have been more propitious. The previous two years had seen the
discovery of the high temperature superconductors
LBCO~\cite{bednorz1986possible} and
YBCO~\cite{wu1987superconductivity}; the discovery of quasi-crystals
had been made 4 years previously~\cite{shechtman1984metallic}; and the
integer~\cite{klitzing1980new} and fractional quantum hall
effects~\cite{tsui1982two} had been discovered and largely explained
during the previous 8 years~\cite{laughlin1983anomalous}.  Five major
discoveries in condensed matter physics in less than a decade,
ultimately garnering 5 Nobel Prizes, and each a splendid manifestation
of quantum mechanics.

Nevertheless, Fisher evidently did not regard his role as to provide a
self-congratulatory pat on the back to condensed matter physics.
Instead, he took it upon himself to explain his thoughts on \lq\lq...
what condensed matter physics does, or should be doing, and what
defines condensed matter physics, and thence to approach the question
\lq Does quantum mechanics matter?\rq ".  The viewpoint that he
expounded was at the time still rather unorthodox in many areas of
physics and science, in particular his perspective that \lq\lq the
question of connecting the models with fundamental principles is {\em
not} a very relevant issue or central enterprise."  This would come as
news to those in the physics community engaged in the reductionist
program, including to some extent condensed matter physicists but to a
far greater extent those working in what at the time was sometimes
called elementary particle physics or fundamental physics, and what is
now generally called high-energy physics, or in some areas, with less
hubris and reflecting the progress in accelerator physics and budget,
medium-energy physics.  This dichotomy within physics was certainly
pervasive at the time, and important enough that Fisher chose to make
it the focus of his article.

Fisher undoubtedly hoped to use the opportunity provided by Feshbach to
pop the bubble of the reductionist approach to condensed matter
physics, and this he did with gusto, and one assumes, with the twinkle
in his eye and suave\cite{suave} cravat by which I will always remember
him.

Over the last 35 years since Fisher's article, physics and its
philosophical roots have undergone a significant shift.  This shift
influences the way we do physics, the way we interpret what we do, and
the way we use physics to explore interdisciplinary scientific
questions. Accordingly, I have chosen for my title a deliberately
ambiguous and evocative phrase. \lq\lq There is plenty of room in the
middle" is of course a parody of Feynman's famous lecture \lq\lq There
is plenty of room at the bottom", which is often credited with
forseeing the nanotechnology revolution \cite{Feynman}.  In my case, I
am drawing attention away from the microscopic scale, to focus instead
on the middle scale or level of description, which depending on the
problem might be the scale intermediate between the lattice spacing and the correlation
length (in the case of critical phenomena) or the scale
larger than the dissipation scale and the scale of energy input (in the
case of fluid turbulence).  In the first of these, one can make this
scale arbitrarily large by tuning the temperature close to the critical
temperature; and in the second, by making the Reynolds number large.
This middle scale was where many of Fisher's interests lay,
understanding the universal, cooperative phenomena that arise there. And of course
that is why he chose his provocative theme for his article.

But I have a second motivation for my title.  On more than one
occasion, I remember Michael saying \lq\lq You can't have
interdisciplinary research without first having disciplines" (for my
personal comments, I will lapse into first name terms). Thus, the
\lq\lq middle" in my title also refers to the space between
disciplines, where interdisciplinary research happens and
transdisciplinary research begins.  I happen to believe that this is
where the greatest scientific excitement can often be generated. Fisher
himself was one of the shining examples of an interdisciplinary
scientist, holding for many years at Cornell University the Horace
White Professorship of Chemistry, Physics, and Mathematics, each of
which he illuminated with his scholarship and unique contributions.
Thus, I propose to use this article to pay tribute to Fisher's deep
understanding of the way in which his most important work on
renormalization group theory led to a transformation in our
understanding of physics and its explanatory purpose, and to extend the
scope of his examples and comments, especially with regard to
non-equilibrium statistical physics. I will spend very little time on
critical phenomena and exponents, because I believe that although this
was the problem that Fisher and the statistical mechanics community set
out to solve, the impact of the renormalization group is equally
important away from criticality.  One of the ways that I will argue
this is through the connections between universality, renormalization
group, levels of description and asymptotics. I also want to
argue that the perspective he espoused is especially relevant today to
applications outside of physics, where there has arguably been less
self-consciousness about the way we choose to construct theories.

\section{Condensed Matter Physics: Does Quantum Mechanics Matter?}

To appreciate Fisher's subversive intentions, it is worth reminding the
reader of the intellectual background to his article. It is no
exaggeration to say that physics underwent a remarkable transformation
sometime around the middle of the 20th century.  During the previous
decades, the reductionist program had been profoundly successful, with
the discovery of the electron, neutron, proton; the theoretical
predictions of the neutrino (by Pauli \cite{pauli1930} in 1930) and the
meson (by Yukawa\cite{yukawa1935interaction} in 1934); the detection of
the former by Cowan, Reines, Harrison, Kruse and McGuire
\cite{cowan1956detection} in 1956, and of the latter in 1936 by
Neddermeyer and Anderson \cite{neddermeyer1937note} --- although they
actually discovered what we now call the muon, a lepton whose mass was
close to Yukawa's prediction for the nuclear force carrier mass; the
actual discovery in 1947 of Yukawa's particle, the pion, in cosmic rays
\cite{lattes1947processes}; and accelerating with the discovery of a
multitude of mesons during the 1950's and the subsequent decades.  Yet,
with the birth of solid state physics around 1940 (reflected by the
establishment of the American Physical Society Division of Solid State
Physics (DSSP) in 1947), the initial focus on one-electron properties
of metals and semi-conductors turned in the 1950's to the collective
properties of matter through many-body theory and the application of
non-relativistic quantum field theory.  In due course, the field became
known as condensed matter physics, reflecting the increasing focus on
all matter, not just solids, and especially on collective phenomena
\cite{cmtvssst} (and marked by the American Physical Society renaming
the Division of Solid State Physics as the Division of Condensed Matter
Physics in 1978).

It quickly became clear that novel macroscopic phenomena transcended
the additive properties of single degrees of freedom, and a renewed
focus on collective properties was seen by many as the defining aspect
of condensed matter physics, particularly through the field's visionary
leadership by P.W. Anderson \cite{anderson1972more,anderson2018basic}.
Indeed, Fisher is very clear where he stands on this point
\cite{fisher1988condensed,fisher2017condensed}: \lq\lq The basic
problem which underlies the subject is to understand the many, varied
manifestations of ordinary matter in its condensed states and to
elucidate the ways in which the properties of the \lq\lq units" affect
the overall, many-variable systems".  In other words, condensed matter
physics is about interactions primarily.

For present purposes, I would argue that the pivotal developments in
the field of condensed matter theory were two iconically quantum
mechanical theories: Bogoliubov's 1947 theory of superfluidity for the
weakly-interacting Bose gas \cite{bogoliubov1947theory}, and the
Bardeen-Cooper-Schrieffer 1957 theory of superconductivity
\cite{bardeen1957microscopic,bardeen1957theory}.  I choose these
problems, because they are relevant to Fisher's conception of the task
of condensed matter physics.  Again, he is very explicit about the
questions that animate the task of addressing the basic problem
underlying condensed matter physics: \lq\lq What are the states of
matter? ... What is their nature? ... How do the various states
transform into one another?"  As we will see, inextricably linked with
these questions are two concepts that are subtle and took a long time to
be appreciated: the notions of \lq\lq minimal models" and \lq\lq
levels of description".   In fact, there are several features of
superfluidity and superconductivity theory to which I especially want
to draw attention.

\subsection{Minimal models}

First, their starting point. Both these works showed the surprising
explanatory power of a ludicrously simple model of interacting Bosons
or electrons respectively that is clearly a brutal idealization of the
real complexities of atomic or electron-phonon interactions. Bogoliubov
assumed that the interactions between Bosons were weak in the sense
that \lq\lq corresponds to a neglection of the finiteness of molecular
radius, since we do not take into account the intensive increase of
[the potential] for small $r$, which causes the impenetrability of
molecules" \cite{bogoliubov1947theory}.  At the end of his paper, he
argued that the approximation could also be extended to include the
finite \lq\lq molecular" radius by replacing the Fourier transform of
the potential by the amplitude of the binary collision probability in
the Born scattering approximation (an observation for which he thanks
Landau).  This amounts to replacing the real interaction with a contact
interaction $U\delta(\textbf{r})$. Bogoliubov's explicit recognition of
the fact that his approximation was valid at long-wavelengths
where the atomic (molecular) size was far smaller than the
mean particle separation was prescient, but it took several decades for
this insight to be translated into a renormalization group description
\cite{kolomeisky1992renormalization,fisher1988dilute,
bijlsma1996renormalization,andersen2004theory}. Bardeen, Cooper and
Schrieffer (BCS) took the Bardeen-Pines Hamiltonian for electron-phonon
interactions including Coulomb effects \cite{bardeen1955electron}, and
replaced the complex interaction with one in which \lq\lq each state is
connected to {\it n} other states by the same matrix element $-V$\rq\rq
\cite{bardeen1957microscopic}.  They explained that this idealization
applied to a subset of states that were paired with equal and opposite
spin and momentum, and that this was sufficient to capture the
formation of the condensed state, for arbitrary weak matrix
element/potential $V$.

The models that both Bogoliubov and BCS used in
their work are examples of what are sometimes known as \lq\lq minimal
models" \cite{oono1985statistical,goldenfeld1992lectures,batterman2001devil,
batterman2002asymptotics,batterman2014minimal}, \lq\lq model[s] that most
economically caricature the essential physics"
\cite{goldenfeld1992lectures}.  However, what this actually means is
rather subtle, and sometimes misunderstood; in order to elaborate, we
need to discuss the role of asymptotics, and in particular the process
by which the minimal models were analyzed.

\subsection{Asymptotics and Universality}

One reason that both Bogoliubov and BCS were successful where others
had failed, was because they used non-perturbative methods to extract their
physical conclusions.  BCS used a variational method, whilst Bogoliubov
invented a canonical transformation applied after the na\"ive
perturbation theory in the weak potential had been performed. Shortly
after the publication of the BCS letter announcing their results
\cite{bardeen1957microscopic}, Bogoliubov's method was applied to the
BCS model \cite{bogoliubov1958new,valatin1958comments}.  I emphasize
the non-perturbative nature of the theories, because in both cases, the
results are non-analytic: for example, the depletion of the condensate
in the Bose gas problem at zero temperature due to interactions is
proportional to $U^{3/2}$, whilst in the BCS problem, the energy gap
depends on the matrix element $V$ and the density of states at the
Fermi level $N(0)$ through an essential singularity of the form
$\exp(-1/N(0)V)$.  Such results are, of course, beyond the scope of
simple, regular, finite order perturbation theory and due to the way
they are obtained, are believed to be valid asymptotically as leading
order approximations for $U,V\rightarrow 0$.

The necessity of using non-perturbative methods to obtain non-analytic
results that describe the emergent states of superfluidity and
superconductivity is not a result of any \lq\lq exotic" character of
these states.  Phase boundaries are by definition the loci where the
partition function is non-analytic, and the emergent states are
physically not simple perturbations around the normal state, possessing
generalized rigidity and elasticity as a result of the symmetry
breaking associated with the transition
\cite{anderson2018basic,goldenfeld1992lectures}. The non-analytic
nature of phase transitions means that any mathematical expansion about
the transition point cannot be convergent, and so our results are at
best asymptotic in the minimal model interaction parameter.   As a
result, there will be sub-leading corrections to our predictions, in
addition to other regular terms that may arise from the details ignored
in constructing the minimal model in the first place.  I also want to
stress that in both the examples given here, the non-analytic results
arise even at mean field level, and proper treatment of the interaction
of fluctuations adds another level of non-analyticity.


For example, in the case of the Bogoliubov Bose gas,
the minimal model itself is constructed to be an asymptotic
long-wavelength limit when the potential range is negligible compared
to the mean separation of the particles.  Corrections to this
approximation arising from including a finite potential range or a
non-vanishing particle density will decorate the leading order
asymptotic result in a complicated way that needs careful consideration
of the various approximations. Nevertheless, the notion of a minimal
model as being an economical caricature of the essential physics must
be understood in this asymptotic context.  The two cannot be separated.
Thus, it is not simply that a minimal model is an inaccurate
representation of a physical system, a poor approximation that can be
improved by embellishment.  Instead the point is that the explanatory
requirement of a minimal model is to account for the phases and phase
transitions of matter, just as Fisher emphasized, and also the
mathematical structure of the description of the latter, which
necessarily involves asymptotic methods.  Gratuitously realistic
details of the starting model will not only complicate the technical
task of extracting the asymptotic structure, but will not impact the
leading order behavior.  Thus, a minimal model represents a
universality class of models; whether or not one uses the minimal model
or a decoration of it, the asymptotic outcome of a non-perturbative
calculation will not be different.

Overall, we are led inexorably to the conclusion that there is a strong
connection between asymptotics and universality.  Indeed, in my own
work, I have shown how renormalization group methods can be used to
treat singular mathematical problems, such as those arising in
differential equation theory
\cite{goldenfeld1992lectures,chen1996renormalization} and low Reynolds
number fluid dynamics \cite{veysey2007simple}.  In these problems,
there are no fluctuations at all: the non-analyticity arises at mean
field level.

\subsection{Minimal models in action: the case of superconductivity}

This narrative of the nature of minimal model explanation is not a
subjective philosophical interpretation that is open to debate,
although it has been challenged \cite{lange2015minimal} and defended
\cite{mckenna2021lange} in the philosophical literature
\cite{batterman2001devil,batterman2014minimal,batterman2021middle}.
It is what physicists actually do. In the case of BCS theory, we have a
very important demonstration of the validity and utility of this
perspective. The BCS minimal model of superconductivity leaves out a
number of embellishments that are desirable to include if one wants to
calculate accurately observables such as the transition temperature or
energy gap at zero temperature for materials where details of the
actual electron-phonon interactions are measurable.  Such an
elaboration of the BCS theory was initiated by Eliashberg
\cite{eliashberg1960interactions}, who accounted for the detailed form
of the electron-phonon frequency spectrum even in the strong coupling
limit, using methods developed by Migdal for the normal state
\cite{migdal1958interaction}, and by McMillan
\cite{mcmillan1968transition} who provided a synthesis that included
electron-electron interactions and band structure effects. In
Eliashberg's theory, the effects of phonons lead to integral equations
accounting for the phonon excitation spectrum and so the theory becomes
intrinsically non-local.  In fact, the interaction between electrons,
mediated by phonons, is inherently a retarded interaction, because of
the mass difference between the lattice ions contributing to the phonon
modes and the electrons themselves.  Distortions of the crystal
lattice, described by the phonons, inevitably relax on a much slower
time scale than that of the electrons, and basically, it is the fact
that electrons experience the local deformations that causes them to
interact and be attracted to one another, in terms of their kinematics
--- hence the BCS coupling of electrons together in time-reversed
momentum states. In McMillan's main result
\cite{mcmillan1968transition} (Eq. (18)), the structure of the BCS
formula, with its essential singularity is modified to $ \exp (-1.04
(1+\lambda)/(\lambda - \mu^*(1+0.62 \lambda))$ where $\lambda$ and
$\mu^*$ are constants that are respectively calculable from the
Eliashberg theory and the effects of Coulomb interactions
\cite{morel1962calculation}.  For small values of the parameter
$\lambda$ (i.e. the weak coupling limit), the formula reduces to the
BCS formula, with $\lambda-\mu^*$ being identified as $N(0) V$.  Thus,
we see explicitly how the minimal model can be embellished by including
greater levels of realism in order to compare with experimental
details, but staying within its universality class and mathematical
structure.

\subsection{Levels of description and emergence}

Bogoliubov's work showed how an interacting system can, within a first
approximation, be treated as a collection of non-interacting
quasi-particles each of which involves all the original \lq\lq real"
atoms in the system.  The quasi-particles were recognized right at the
outset as forming an explicitly long-wavelength description of the
system, the details of which were later developed in 1957 by Lee and
Yang, and others
\cite{lee1957many,lee1957eigenvalues,brueckner1957bose,
kolomeisky1992renormalization,fisher1988dilute,
bijlsma1996renormalization,andersen2004theory}.  The many-body ground
state and the excitations of the interacting Bose system depend on a
single parameter $U$ characterizing the strength of interactions of the
original, \lq\lq real" Bosons; both the ground state and the
excitations are collective properties of the system.  In the case of
superconductivity, the additional complication is the formation of
composite bosons --- Cooper pairs --- which simultaneously undergo a
similar ordering as in the Bogoliubov gas.  The upshot of these
developments is that for many purposes, a superfluid or superconductor
has a universal description that effectively hides certain details of
the underlying constituents and their interactions. In the case of the
weakly interacting Bosons, the interaction parameter $U$ depends on the
s-wave scattering length derived from the collision of two Bosons, and
the theory is valid when this length is much shorter than the mean
separation of the Bosons.  Thus, it can be said that $U$ relates two
different {\em levels of description}, the microscopic world inhabited
by the minimal model of interacting Bosons, and the macroscopic world
of the condensate with its superfluid properties and excitation
spectrum for the quasi-particles that have emerged.  It would be
possible in principle to infer the existence of the quasi-particles
phenomenologically from thermodynamic measurements, as Landau famously
did \cite{landau1941two}, without knowledge of the properties or even
the existence of the atomic level of description.

Although much of the impact of the Bogoliubov and especially the BCS
theories derived from their ability to bridge levels of description and
compare the predictions of minimal models quantitatively with
experiment at the microscopic level of descriptions (e.g. the work of
McMillan \cite{mcmillan1968transition} already mentioned), Fisher is
emphatic that this is not their main importance. We already cited his
views about connecting \lq\lq the models with fundamental principles",
and elsewhere in the same article he writes \lq\lq The basic problem
which underlies the subject is to understand the many, varied
manifestations of ordinary matter in its condensed states ... an
important role is, and, indeed, should be played by various special
`models'.  ... One might as a theorist argue that more attention should
be paid to the connection between models and the fundamental principles
as embodied in quantum mechanics."
\cite{fisher1988condensed,fisher2017condensed}  However, he argues,
this is misguided for the following reason: \lq\lq I stress again, that
in any science worth the name, the important point is to gain
understanding. The language in which the understanding is best
expressed cannot be dictated ahead of time, but must, rather, be
determined by the subject as it develops. Accordingly, it really would
not be a \lq great success' to derive the Ising model from atomic
theory or quantum mechanics. Indeed, for which physical system would
one be `deriving' it?" \cite{fisher1988condensed,fisher2017condensed}
His point is that a minimal model such as the Ising model is a minimal
model for a ferromagnet, a liquid-gas systems, an alloy, a
ferroelectric material and, I would add, many other systems, including
examples in biology and ecology ranging from the neurobiology of
retinal ganglion cells \cite{cocco2009neuronal} to the annual nut
distribution of pistachio trees \cite{noble2018spatial}.  Thus,
although \lq\lq ... one does want to understand what the important
variables actually are ... in many ways, it is more important to
understand how these variables interact together and what their
`cooperative' results will be".
\cite{fisher1988condensed,fisher2017condensed}

To close this important section on levels of description, I want to
comment again, in Fisher's prescient words, of the important role
played by asymptotics: \lq\lq Personally, I find the elucidation of the
connections between various disciplinary levels a topic of great
interest. Typically, delicate matters of understanding appropriate
asymptotic limits, both physical and mathematical, are entailed. The
issues involved are often very subtle; nevertheless, one must admit,
they seldom add significantly to one's understanding of either the
`fundamental' starting theory or of the target discipline. Sad but
true!" \cite{fisher1988condensed,fisher2017condensed}.

To this I would add a point that is not made frequently enough:
technical derivations of higher levels of description from more
microscopic ones have a limited regime of validity. For example, it is
possible to derive the Navier-Stokes equations from Boltzmann's kinetic
equation.  This is a long and complicated calculation, that I often
show the physics students in my statistical mechanics classes.  After I
have done that, I show them a phenomenological motivation that of
course is very brief and closer in spirit to the derivations usually
given in fluid mechanics texts.  Most students, when asked, find the
former derivation more convincing.  They are surprised to learn that I
do not share this view, and are somewhat shocked to realize that the
seemingly more careful, technical derivation is only valid at very low density,
whereas we know that the Navier-Stokes equations are an excellent
description of fluid phenomena up to very high densities.

\subsection{Universal and asymptotic scaling laws in non-critical matter}

Fisher structured his article by first describing his perspective on
the nature of condensed matter physics, and secondly by giving multiple
examples of forms of matter to illustrate his main thesis.  The first
form of condensed matter that he chose was polymeric matter, focusing
first of all on the excluded volume problem for a single polymer
\cite{flory1949configuration,edwards1965statistical}, and then the
problem of the statistical thermodynamics of dilute polymer solutions
\cite{ohta1982conformation,oono1985statistical}.  Fisher chose this
example and not the obvious example of critical phenomena scaling. Why?

I conjecture that he wanted to make two points.  First, that there are
other ways in which matter can enter asymptotic realms other than just
being close to a critical point. Fisher was intimately acquainted with
asymptotic methods and understood that an example of scaling which has
nothing to do with a critical point would be consciousness-raising for
some of his audience.
%
%
This is elaborated on with numerous ``critical phase'' examples in the
chapter of this volume by Leo Radzihovsky.  Second, that he wanted to
show the relevance of condensed matter physics to a discipline other
than physics.  He writes
\cite{fisher1988condensed,fisher2017condensed} \lq\lq ... other
challenging problems remain beyond the size of a single isolated
molecule.  One might say, `Well, aren't polymers for chemists?'  The
answer is that many of their properties have proved too difficult for
traditionally trained chemists!"

At this time, Fisher was in fact a Professor of Chemistry --- amongst
other things --- and had in the early part of his career at King's
College, London, worked extensively on the statistics of self-avoiding
walks and Ising models, developing methods to enumerate configurations,
extract exponents using series expansions, Pad\'e approximants and so
on.  In an especially interesting paper from this period
\cite{fisher1964ising}, he and Gaunt estimated the behavior of
self-avoiding walks and Ising models in dimensions $d > 4$, finding
that the deviations from mean field theory seemed to become very small
above four dimensions, a result that must have been on his mind during
the development with Wilson of the $\epsilon = 4-d$ expansion
\cite{wilson1972critical}.  One of the earliest applications of
renormalization group theory, and in particular the
$\epsilon$-expansion, was to the conformations of polymers.

In 1965, Edwards had \cite{edwards1965statistical} developed a field
theoretic approach to the statistics of a self-avoiding polymer chain,
based on Bogoliubov's idea of the pseudo-potential
\cite{bogoliubov1947theory}, and used a functional integral approach to
derive a self-consistent theory that showed how in the limit of an
asymptotically long chain, the radius of gyration of a single polymer
scaled with length $L$ as $L^{3/5}$.  This theory, like Bogoliubov's,
had a regime of validity associated with a description at long
wavelengths compared to the range of interparticle forces, in this case
being that $L$ was much greater than the monomer scale.  Edwards' result, is
like Bogoliubov's, a mean field theory approximation of course, and ignores fluctuations which
add still another level of non-analyticity to the problem.  The much more
complicated problem of semi-dilute polymer solutions was treated the
following year by the same methods \cite{edwards1966theory} subject to
the same limitations of course.


In the early 1980's, Ohta and Oono
\cite{ohta1982conformation,oono1985statistical} developed a
conformation-space renormalization group method which was not based on
de Gennes' analogy to the $n$-component magnet \cite{de1972exponents}
and thus was able to account for arbitrary molecular weight
distributions of the polymers.  Ohta and Oono used their method to go
far beyond computing simple scaling laws, not only correcting the Edwards mean field
results to include fluctuations, but also calculating
the universal scaling functions for the statistical thermodynamics of
polymer solutions as a function of concentration.  Fisher could not
hide his excitement in reporting their results, especially the complete
comparison with experiment \cite{wiltzius1984dynamics} over several
decades for the functional form of the osmotic compressibility and
other variables.  In particular, Fisher enthused: \lq\lq Thus we see in
polymeric matter new, subtle and universal behavior which we have
succeeded in understanding theoretically. But quantum mechanics has had
essentially nothing to say about the problem! Indeed, one feels that if
some of the giants of the past, like Boltzmann or Gibbs or Rayleigh,
were able to rejoin us today, they would be able to engage in research
at the cutting edges of condensed matter physics without taking time
off to study quantum mechanics first!"
\cite{fisher1988condensed,fisher2017condensed}

\section{Turbulence: does fluid mechanics matter?}

I want to turn now to a topic that did not play a large role in
Fisher's article or his research interests: systems far from
equilibrium.  At the time of his article, condensed matter physicists
were beginning to address problems in non-equilibrium pattern
formation, phase transition kinetics, and there is a long story to tell
there about asymptotic scaling laws, since this is the field where I
started my own career.  That story is for a future occasion.  Here, I
want to mention briefly two vignettes from my own research on turbulent
fluids.  The first of these I know he enjoyed, because I got to tell
him about it during a visit to Cornell.  Unfortunately, I do not
remember if I had the chance to tell him about the second one.

\subsection{A turbulent analogue of Widom scaling}

The classic problem of fluid dynamics is that fluids appear to be scale
invariant when strongly turbulent
\cite{pope2000tf,sreenivasan1997phenomenology}.  Specifically, in a
fluid geometry with characteristic scale $D$ (which might be the
diameter of a pipe), characteristic velocity $U$ (which might be the
average mean flow velocity along a pipe) and kinematic viscosity $\nu$,
we define the Reynolds number as $Re = UD/\nu$.  In addition, there is
the energy dissipation rate $\epsilon$, which is in steady state equal
to the energy input rate.  The so-called energy spectrum --- the
kinetic energy per unit mass of fluid per unit wavenumber $k$ --- was
argued by Kolmogorov \cite{KOLM41}, in a paper universally known as
K41, to vary as $E(k) = \epsilon^{2/3} k^{-5/3}$ at large $Re$, and
this is broadly in agreement with experimental data going back more
than 50 years \cite{GRAN62}.  The classical measurements and theories
only concerned themselves with the lowest order moments of the velocity
fluctuation probability density, but today there is strong evidence for
multifractal scaling of higher order moments
\cite{sreenivasan1997phenomenology}.  Kolmogorov further showed that
the range in $k$-space over which this scaling occurred was
intermediate between the scale of forcing $D$ and the Kolmogorov scale
of dissipation $\eta_K= (\nu^3/\epsilon)^{1/4}$, in other words $2\pi/D
\ll k \ll 2\pi/\eta_K$.  As $Re\rightarrow \infty$, this range of
scaling increases because the Kolmogorov scale gets smaller and
smaller, but can be shown to always be above the mean free path, so
that the fluid is always in the hydrodynamic limit.  In fact, K41 is
not quite correct, even for the second order correlation function, and
there is a so-called large-scale intermittency correction, $\eta$,
which changes the scaling result to $E(k) \sim k^{-5/3 + \eta}$
\cite{kolmogorov1962refinement,sreenivasan1993update}.


To a statistical mechanic, this scaling behavior is reminiscent of what
happens in a magnet near a critical point.  In general, correlation
functions decay as a power-law within a range of wavenumber that is
intermediate between the lattice spacing $a$ and the correlation length
$\xi$.  As the temperature $T$ approaches its critical value $T_c$, the
correlation length diverges to infinity, and at any scale accessible to
experiment, the correlations will be in the power-law scaling range,
ultimately scaling as $G(k) \sim k^{-2 + \eta}$ at the critical
temperature itself.  Here $\eta$ is a correction to the mean field
result, sometimes called Fisher's exponent, and now understood to be
the anomalous scaling dimension of the magnetization $M$, which grows with
a power-law $\beta$ as $T\rightarrow T_c^-$.


Statistical mechanicians also know that these results are only valid
when there is no external field $H$.  When $H\neq 0$ there are more
complicated scaling laws.  For example, in general we expect that the
magnetization is proportional to the external field, but this linear
response law breaks down exactly at the critical temperature: $M (H,
T_c) \sim H^{1/\delta}$, where $\delta$ is another critical exponent
that can of course be calculated by renormalization group.  In the
critical region $H \sim 0$ and $t\rightarrow 0$ (where $t\equiv
(T-T_c)/T_c$), Widom showed \cite{WIDO65} that the behavior of the
magnetization, ostensibly a function of both $H$ and $t$ is actually a
function of a combined variable, predicting a data collapse of $M(H,T)$
when plotted appropriately, similar to the data collapse that we talked
about with polymer solution theory.  This scaling was independently
discovered by Kadanoff, and explained in a famous paper the following
year \cite{kadanoff1966scaling}, and this discovery by Widom and
Kadanoff was instrumental in the development of the renormalization
group.

Thus, it is a natural question to ask: is there a counterpart to the
Widom scaling in turbulence?



I approached this question by asking what are the analogous quantities
to $t$ and $H$ in turbulence.  With Greg Eyink, I had long ago argued
that it is natural for $1/t$ to be analogous to $Re$, because both
control the size of the intermediate regime in $k$ space where there is
power-law scaling as $1/t$ and $Re$ respectively go to infinity
\cite{EYIN94}. But what is the turbulent analogue of $H$?  My reasoning
was that $H$ is a variable that couples to and induces magnetization.
In a pipe, it turns out that the laminar flow is linearly stable, but
if it has rough walls, the roughness will excite turbulence.  Thus, my
guess was that the roughness scale $s$ could be analogous to $H$.
Using the same sort of scaling arguments that Kadanoff and Widom had
used, and assuming that turbulence was in fact a non-equilibrium steady
state with its own critical point at $Re \rightarrow \infty$,
$s\rightarrow 0$, it was possible to predict the analogue of Widom's
scaling law \cite{GOLD06}.

Fortunately, there were experimental data on this very question!  In
1933, Nikuradse had measured the friction experienced by a turbulent
fluid as it transited a pipe with rough walls \cite{NIKU33}.
Importantly, Nikuradse had systematically varied both $Re$ and the
roughness scale $s$.  His data collapsed satisfactorily when plotted
according to the formula \cite{GOLD06}.  However, I had only used the
mean field exponents in my collapse, the ones found in K41.  Mehrafarin
and Pourtolami extended my calculation to include the intermittency
correction \cite{MEHR08}.  Even though the correction is very small for
the second order correlation function, they were able to show
convincingly that they could extract it by improving the data collapse
fit.  Their result was consistent with known estimates for the
intermittency exponent, obtained by direct measurements of velocity
fluctuations.

This result is extraordinary for the following reason.  In phase
transition theory, it is known that there is a connection between
purely thermodynamic critical exponents and those associated with
spatial correlations \cite{goldenfeld1992lectures}.  It turns out that
it is possible to deduce Fisher's exponent for correlations, the
anomalous dimension of the order parameter, directly from the
thermodynamic exponent $\delta$ that we mentioned above quantifies the
breakdown of linear response theory at the critical point.  In my
treatment of the turbulent scaling problem, the analogue of the
thermodynamic exponent is one that concerns the scaling of the pressure
drop along the pipe with $Re$.  So, using the data collapse scaling
law, if it had been known in 1933, Nikuradse could have measured
indirectly the intermittency exponent of turbulence characterizing the
strong spatial fluctuations!!  In other words, the fluctuations of
turbulence are directly related to the pressure drop along a pipe, i.e.
the dissipation experienced.  This implies that there are
non-equilibrium fluctuation-dissipation relations in turbulence, but at
the present time, we do not know how to extract them.

There were other developments that are important to mention.  My
Illinois colleagues Pinaki Chakraborty and Gustavo Gioia came up with
an ingenious heuristic mean field argument that predicted the various
exponents that are visible in different regimes of the Nikuradse data
\cite{blasius1913ahnlichkeitsgesetz,STRI23}, at least at mean field
level \cite{GIOIA06}.  With Nicholas Guttenberg I worked out the
scaling laws and simulated the flow in a rough two-dimensional pipe
(i.e. a soap film suspended between two wires) \cite{GUTT09}, where the
point was that in two-dimensions, a different scaling of the velocity
correlations is possible than the K41, because of the flow of angular
momentum fluctuations (enstrophy).  This fact meant that we could in
principle construct flows with different velocity correlation scaling
from the K41 one, and observe the effect on the pressure drop, i.e. the
friction, and thus test the fluctuation-dissipation relation that had
been discovered.  Our predictions were fully confirmed in experiments
that we performed with Walter Goldburg at Pittsburgh and Hamid Kellay
in Bordeaux \cite{TRAN10,kellay2012testing,vilquin2021asymptotic}.

Overall, these findings suggest that fully-developed turbulence is
controlled by a non-equilibrium critical point, with strong connections
to a far from equilibrium statistical mechanics through an unknown
fluctuation-dissipation relation.

However, this was not the only surprise in treating turbulence as a
problem in statistical mechanics \cite{goldenfeld2017turbulence}.


\subsection{Fluids become turbulent through a non-equilibrium phase
transition}

Up to now, we have talked about the large $Re$ behavior of turbulence.
But how do fluids transition from predictable, smooth, laminar flows to
unpredictable, fluctuating turbulent flows?  In certain shear flows
where the laminar-turbulence transition is sub-critical, including pipe
flow, we now have compelling theoretical
\cite{pomeau,barkley2011simplifying,goldenfeld2010extreme,sipos2011directed,ppmodel,
barkley2016theoretical,wang2022stochastic}, computational
\cite{chantry2017universal,gome2020statistical} and experimental
\cite{hof_dp_1d_couette,mukund2018critical,klotz2022phase,Lemoult2023}
evidence that this transition is a non-equilibrium phase transition in
the universality class of directed percolation
\cite{hinrichsen2000non}.  A synthesis is beginning to emerge
\cite{goldenfeld2017turbulence,hof2023directed,avila2023transition},
and most workers in this small field tend to agree that the
one-dimensional problem is understood in varying levels of detail, but
the problem of the transition to turbulence in two-dimensions is still
not understood theoretically, even though the experimental data are
very compelling \cite{klotz2022phase}.

I believe that Fisher would have been excited by these developments,
because it is fascinating to see phase transition physics, which he
calls \lq\lq my own love", emerge unexpectedly in a field such as fluid
dynamics, where there is no explicit stochasticity, no partition
function, and no obvious connection to directed percolation.  This is
an intriguing example of universality, to be sure.  But I think Fisher
would have been equally delighted by the way that this prediction was
made.  When my students and I set out to tackle this problem, our
perspective was that the worst starting point for understanding the
transition to turbulence was the Navier-Stokes equations.  Instead, our
goal was to construct the level of description that corresponded to
what would be Landau theory for an equilibrium transition.  To this
end, we used direct numerical simulations to identify the weak
long-wavelength modes that are relevant at a putative continuous
laminar-turbulence transition, and once we had identified these,
constructed the generic description, which turned out to be a
stochastic predator-prey or activator-inhibitor model \cite{ppmodel}.
In other words, we wanted to identify the variables that defined the
universality class of the laminar-turbulence transition. This model led
to directed percolation in a straightforward way using known techniques
\cite{mobilia2007}, and we showed was able to reproduce the universal
aspects of the phenomenology at the laminar-turbulence transition
\cite{ppmodel,wang2022stochastic,goldenfeld2017turbulence}.

We did not set out to derive the directed percolation transition from
the Navier-Stokes equations, because we already knew that this sort of
derivation from low levels of description cannot be done
systematically, let alone rigorously, even for the simplest systems
such as equilibrium magnets.  We also did not attempt to predict the
critical Reynolds number of the laminar-turbulence transition, because
we know from the renormalization group in equilibrium statistical
mechanics that this is not universal.

In a later paper reviewing the statistical mechanics approach to
turbulence \cite{goldenfeld2017turbulence}, we did give a heuristic
motivation for the predator-prey description, at least at mean field
level, but I believe that it is fair to say that the majority of fluid
mechanicians do not yet share Fisher's perspective that it is more
important to understand the cooperative interactions of the emergent
degrees of freedom at the appropriate level of description than to
systematically derive these degrees of freedom.  The reason for his
perspective, and the reason why it is so counterintuitive, is that our
algorithm is to heuristically guess the emergent level of description,
and then use the renormalization group picture to identify only the
relevant degrees of freedom at the appropriate fixed points, thus
obviating the need to calculate quantities that will not affect the
final goal of our understanding.

\section{Epilogue}

It is an unavoidable temptation to imagine the reaction when Fisher's
manuscript arrived on Feshbach's desk.  Not only does its very title
threaten to undermine the rationale for the whole enterprise, but
Fisher doubles down, arguing at one point that \lq\lq \dots it is a
mistake to view complementarity as merely a two-terminal black box
\dots a fully reductionist philosophy, while tenable purely as
philosophy, is the wrong way to practice real science!"  For many
editors and organizers of a symposium on the legacy of Niels Bohr,
Fisher's article might have seemed like an unwarranted provocation. But
I suspect that in Feshbach's case, he  appreciated the article as a
masterly contribution that both recalled the extraordinary collision
between physics and philosophy of a previous age, and heralded the dawn
of a new era where this confrontation would be renewed, but on fresh
ground, whose boundaries had been demarcated by Fisher's warning shot
across the bow.

Surprisingly, it would be over a decade later before the first
skirmishes took place. I would place their date to be the 2001
publication of a book by R.W. Batterman, entitled \lq\lq The devil in
the details: Asymptotic reasoning in explanation, reduction, and
emergence" \cite{batterman2001devil}.  Drawing on sources such as the
works of Michael Berry on caustics and asymptotics
\cite{berry1990beyond,berry1995asymptotics}, Leo Kadanoff
\cite{kadanoff1966scaling}, Michael Fisher \cite{fisher1983scaling} and
Ken Wilson \cite{wilson1974renormalization} on critical phenomena and
renormalization group, and my own work on asymptotics and
renormalization group
\cite{goldenfeld1989intermediate,goldenfeld1990anomalous}, as
summarized especially in \cite{goldenfeld1992lectures}, Batterman
weaves together the concepts of minimal models, asymptotics,
renormalization group theory and emergence in a thesis that explicitly
recognizes how the modern way of doing science is neither Popperian nor
Kuhnian, partly because of a more sophisticated notion of
falsifiability entailed by concepts such as universality.  This work
and its extensions \cite{batterman2014minimal,batterman2021middle} have
been increasingly influential during the last 20 years, and the battles
have been fought mostly by philosophers of science. Most recently,
Batterman has used hydrodynamic linear response theory as an example to
expound further on these ideas, echoing many of the themes that Fisher
advocated and which I have described here \cite{batterman2021middle}.

I believe that the philosophical debates about minimal models are
especially important for complex systems such as biology and turbulent
fluid mechanics, because the choices one makes in modeling are so much
more difficult than in physics.  I discussed above how the minimal
model and renormalization group approach to modeling at the appropriate
level of description was successful in understanding quantitatively the
laminar-turbulence transition in certain shear flows.  The idea of
making effective models of turbulence is, however, not new to fluid
mechanics.  Indeed, there is a long tradition of making approximate
models, and these are discussed in the textbooks, especially Pope's
excellent monograph on turbulent flow, section 8.3 \cite{pope2000tf}.
These are not minimal models, because the \lq\lq ultimate objective is
to obtain a tractable quantitative theory or model that can be used to
calculate quantities of interest and practical relevance."
\cite{pope2000tf}  This means using models as approximations for
computer simulations or simple analytical calculations whose utility is
assessed by factors such as cost and ease of use, range of
applicability and accuracy \cite{pope2000tf}. In general, it is not the
goal to devise universal scaling functions, for example, in the way
that minimal modeling was used in the semi-dilute polymer problem
\cite{ohta1982conformation,oono1985statistical}.

That being said, it must not be forgotten that Kolmogorov's program of
analyzing turbulence began with two similarity hypotheses about universality, and this is
to my knowledge the first example of where the renormalization group
perspective was articulated clearly \cite{KOLM41}, and followed up in
many seminal works by Barenblatt (Kolmogorov's last student!)
\cite{barenblatt1972self,barenblatt96scaling}.  The first of these is
that the energy spectrum can be written in the middle or inertial range
of scales as $E(k) = (\nu^2/\eta_K) F(k\eta_K)$, where the limit
$kL\rightarrow\infty$ has been assumed to exist and been taken, and the
scaling function $F(z)$ \lq\lq must be the same for all cases of locally isotropic
turbulence" \cite{KOLM41}.  This hypothesis does {\em not} require the large Reynolds
number limit to have been taken, and indeed it has been verified to a good approximation even
for transitional turbulent flows \cite{cerbus2020small}.  Only when a
second similarity hypothesis is taken, that the $Re\rightarrow \infty$
limit exists, does the K41 scaling emerge.  In fact, the connection
with critical phenomena is very close: the first similarity hypothesis
is not quite correct.  The $kL\rightarrow\infty$ limit does not exist,
and the asymptotics exhibit incomplete similarity
\cite{barenblatt1972self,barenblatt96scaling} leading to the existence
of a new exponent, the intermittency exponent that we described
earlier.

In the renormalization group-informed perspective presented throughout
this article, minimal models are in the universality class of the
transition, and thus give a quantitatively accurate account of the
transition. They are not approximations in the sense that they have
inadequate realism. As we have discussed above, the predictions of
minimal models are quantitatively accurate, despite the fact that the
minimal models are seemingly lacking in realism. In biology, there are
many more levels of description than in physical systems, because of
the intertwining of molecular sequence, three-dimensional structure,
large-scale molecular motions, elasticity, gene expression, metabolism
and energy flow, regulation, signaling, cell division, tissue
mechanics, electrical activity, ..., all the way up to ecology and
evolutionary processes.  Choosing the right questions and the right
levels of description is critically important, because there is no
other way forward.  As Fisher wrote: \lq\lq Schrodinger's equation
really does have {\em almost no} relevance to what one actually does in
vast areas of chemistry."
\cite{fisher1988condensed,fisher2017condensed}  I believe that the
analysis of biological modeling from the perspective of philosophy of
science would be a very interesting project, one which might bring a
broader range of topics into the field, and which might ultimately help
biophysicists and biologists in their efforts to understand these
complex systems by being aware of the necessity to choose the right
level of description.  In my own work, there are a number of examples
where this has been attempted, in topics ranging from the evolution of
the genetic code \cite{vetsigian2006collective}, the open-ended growth
of biological complexity \cite{guttenberg2008cascade}, to the
topological scaling laws that have recently been uncovered in
phylogenetic trees \cite{xue2020scale}.


For physicists, it was all over by the early 1990s, as even elementary
particle physicists came to the clear realization that their subject
was about effective field theory, and not about fundamental physical
law. This was a remarkable 180 degree turn for a community which by and
large had regarded renormalizability as a necessary but not sufficient
condition for fundamental quantum field theories, and which still
adhered to the program of first understanding the basic constituents of
matter, so that they could then be combined to understand the low
energy world to which we have direct access.  But within a few years of
Fisher's article, this view was superseded by the general acceptance of
effective field theory, the fascinating history of which has been
beautifully described from a personal perspective by Weinberg
\cite{weinberg2016effective} and summarized intellectually by Georgi
\cite{georgi1993effective}.  I want to end by briefly talking about
effective field theory, through the lens of Georgi's review, because it
is truly remarkable to see the convergence between the modern
perspective of condensed matter, as championed by Fisher, and the
modern perspective of high energy physics, as embodied in effective
field theory.

\subsection{High energy physics: does high energy matter?}


The starting point of effective field theory is the recognition that
the goal has changed.  The goal is not to create a theory of everything
but, according to Georgi, \lq\lq to isolate a set of phenomena from all
the rest, so that we can describe it without having to understand
everything... Fortunately this is often possible. We can divide the
parameter space of the world into different regions, in each of which
there is a different appropriate description of the important physics."
\cite{georgi1993effective}  This is the pragmatic rationale for doing
effective theory, but historically, this is not how it developed. If we
had a theory of everything, it would be awfully unwieldy to calculate
something specific, and so for convenience you \lq\lq use the effective
theory ... it makes calculations easier, because you are forced to
concentrate on the important physics". \cite{georgi1993effective} In
practice this means doing an expansion in scale, shrinking to zero size
the features of the physics that are smaller than the scale of
interest.  As Georgi writes\cite{georgi1993effective}: \lq\lq this
gives a useful and simple picture of the important physics.  The finite
size effects that you have ignored are all small and can be included as
perturbations."  The attentive reader will have noticed that this
passage is uncannily reminiscent of the modeling strategy articulated
earlier in this article by Bogoliubov in 1947.
\cite{bogoliubov1947theory}  In high energy physics, this strategy is
equivalent to removing massive particles, but the price paid for this
massive oversimplification (sorry, I could not resist!) is that the
ultraviolet regularization is now non-trivial, because of the way that
coupling constants vary with scale. Integrating out the high energy
degrees of freedom results in a non-local theory because of virtual
exchange processes with the neglected massive particles.  The program
of effective theory replaces these non-local interactions with local
interactions, that are \cite{georgi1993effective} \lq\lq ...
constructed to give the same physics at low energies ... Thus the
domain of utility of an effective theory is necessarily bounded from
above in energy scale."  This is of course reminiscent of Bogoliubov
(prompted by Landau) connecting his minimal model to reality by
replacing his point interaction amplitude with the amplitude of the
binary collision probability in the Born scattering approximation, and
of course to the Eliashberg extension of the BCS model of
superconductivity.

I pause to note another important connection to condensed matter physics.  The
scale-dependence of interactions was discovered in the context of
critical phenomena by Kadanoff \cite{kadanoff1966scaling} and
Patashinski and Pokrovsky \cite{patashinskii1966behavior} in 1966 but
identified earlier during the development of the renormalization group
\cite{stueckelberg1953normalisation,gell1954quantum}, and used by the
Landau \cite{landau1965collected} school in the context of quantum electrodynamics and the famous
\lq\lq Moscow Zero"
\cite{landau1954removal,landau1954asymptoticelectron,
landau1954asymptoticphoton,landau1954electronmass,landau1955quantum}
--- the divergence of the coupling constant that renders the field
theory ill-defined at high energies.  I refer the interested reader to
a thoughtful modern perspective on the Moscow Zero in the context of
condensed matter physics \cite{jian2020landau} and the remarkable
experimental observation of the scale-dependence of the coupling
constant in graphene \cite{reed2010effective}.

The effective field theory is also bounded below in energy.  This is
because the effective field theory will generate its own massive
particles, and on energy scales smaller than these masses, these
particles can again be eliminated to generate a new effective theory at
a lower energy scale too.  Thus an effective field theory lives in an
intermediate or \lq\lq middle" scale of energy.

Effective field theory is in practice looked at in a different way,
because we do not have any information about the high energy theory
that we invoked above.  So effective field theory is used to describe
the physics of an energy scale of interest up to a certain level of
accuracy, with a small or finite number of parameters that \lq\lq
parameterize our ignorance in a useful way" \cite{georgi1993effective}.
This is not the same as the old-fashioned view of renormalizability in
field theory, because it is expected that with increasing energy, the
non-renormalizable interactions get replaced with a new effective
theory.

Weinberg recounts \cite{weinberg2016effective} how his perspective
started to change in 1976, when he learned about Wilson's approach to
critical phenomena by integrating out the ultraviolet degrees of
freedom, and using the renormalization group equation to ensure that
physical quantities are cutoff independent.  He realised that this
entails introducing \lq\lq every possible interactions, renormalizable
or not, to keep physics strictly cutoff independent.  From this point
of view, it doesn't make much difference whether the underlying theory
is renormalizable or not ... Non-renormalizable theories, I realized, are
just as renormalizable as renormalizable theories" \cite{weinberg2016effective}.

Georgi describes this change in perspective in a remarkable comment
that sounds the death nell for the fundamental viewpoint of elementary
particle physics \cite{georgi1993effective}: \lq\lq How does this
process end?  It is possible, I suppose, that at some very large energy
scale ... the theory is simply renormalizable in the old sense.  This
seems unlikely ...  It may even be possible that there is no end,
simply more and more scales as one goes to higher and higher energy.
Who knows?  Who cares?"

Michael Fisher would agree.



\begin{acknowledgments}

First and foremost I want to acknowledge the many inspirations,
insights and perspectives that I learned from Michael's writings and
personal discussions in our scientific interactions. One of my own
struggles with renormalization group was to figure out how to separate
it from the context of statistical and quantum field theory. The way
Michael presented his own work, as well as his pedagogical
introductions to the field, aided my understanding of the field that he
had helped invent. Indeed, Michael cared deeply about the presentation
of science, as well as its content; and in a pre-powerpoint world, he
was ahead of the curve in bringing his talks to life by writing on his
slides in real time, filling in the gaps he had purposefully left.  In
this way, he managed to combine the immediacy of a blackboard talk with
the opportunity to create a skillful layout that aided the
presentation.  I felt that I had got to know him properly when he
shared with me his secret: the slides were written in indelible marker,
but the in-lecture annotations were written in water-soluble marker,
and were easily washed away after the talk, ready for the next one.  To
the extent that I am successful as an educator and communicator of
science, it is partly because of Michael's example, along with those of
Jim Langer and Sir James Lighthill.

I also want to use this opportunity to acknowledge with thanks my
friend and collaborator Yoshi Oono, whose brilliant and sometimes
iconoclastic perspectives uniquely influenced my understanding of
science, even going back to the time when I was a graduate student.  It
is no accident that Michael had the breadth and intellectual taste to
chose to begin the \lq\lq forms of matter" section of his article with
Yoshi's largely overlooked work on the conformational space
renormalization group of polymeric matter.  I remember that Yoshi was
excited that Michael had highlighted his work in his article, and that
is how I learned of Michael's article.

I also wish to thank my other collaborators and students, whose work
was mentioned here: Gregory Eyink, Hong-Yan Shih, Tsung-Lin Hsieh, Chi
Xue, Zhiru Liu, Xueying Wang, Lin-Yuan Chen, Nicholas Guttenberg, Tuan
Tran, Alisia Prescott, Pinak Chakraborty, Hamid Kellay, Gustavo Gioia,
Walter Goldburg, Maksim Sipos, Olivier Martin, Fong Liu, Kalin
Vetsigian, Carl Woese, John Veysey.

I wish to thank Robert Batterman for many discussions on the philosophy
of science, and for his skillful articulation and observations about
the way modern statistical physicists approach the task of practicing
science.

Lastly I thank Amnon Aharony and Leo Radzhihovsky for inviting me to
contribute this essay, for their helpful suggestions that improved the
manuscript, and for their patience during the unforseen delays.

This work was partially supported by the Simons Foundation through
Targeted Grant ``Revisiting the Turbulence Problem Using Statistical
Mechanics" (Grant No 662985(N.G.)).

\end{acknowledgments}

\bibliography{Fisher.bib}
\end{document}